\documentclass[pre,aps,preprint,showpacs]{revtex4}
\usepackage{epsfig,graphics}
\newcommand{\bea}{\begin{eqnarray}}
\newcommand{\eea}{\end{eqnarray}}
\newcommand{\be}{\begin{equation}}
\newcommand{\ee}{\end{equation}}
\newcommand{\nn}{\nonumber}
\newcommand{\btof}{\gamma_{b \rightarrow f}}
\newcommand{\ftob}{\gamma_{f \rightarrow b}}
\newcommand{\expo}{\frac{1}{2}(\btof - \sqrt{\btof^2+4\ftob})}
\newcommand{\fa}{\frac{1}{2} - \frac{\btof}{2 \sqrt{\btof^2 + 4\ftob}}}
\newcommand{\ta}{\sqrt{\btof^2 + 4\ftob}~r}

\begin{document}
\title{Asters, Spirals and Vortices in Mixtures of Motors and 
Microtubules: The Effects of Confining Geometries on Pattern Formation}
\author{Sumithra Sankararaman}
\email[Email:~]{sumithra@uic.edu}
\altaffiliation{
Present address: Department of Physics, University of
Illinois at Chicago, 845 W Taylor St., Chicago, IL 60607-7059, U.S.A 
}
\affiliation{The Institute of Mathematical Sciences,\\
C.I.T. Campus, Taramani, Chennai 600 113,\\
India.}
\author{Gautam I. Menon}
\email[Email:~]{menon@imsc.res.in}
\affiliation{The Institute of Mathematical Sciences,\\
C.I.T. Campus, Taramani, Chennai 600 113,\\
India.}
\author{P.B. Sunil Kumar}
\email[Email:~]{sunil@physics.iitm.ac.in}
\affiliation{Department of Physics,\\
Indian Institute of Technology Madras, \\
Chennai 600 036,\\
India.}

\begin{abstract}
We model the effects of confinement on the stable
self-organized patterns obtained in the
non-equilibrium steady states of mixtures of
molecular motors and microtubules. In experiments
[N\'ed\'elec {\it et al.} Nature, {\bf 389}, 305
(1997); Surrey {\it et al.}, Science, {\bf 292},
1167 (2001)] performed in a quasi-two-dimensional
confined geometry, microtubules are oriented by
complexes of motor proteins.  This interaction
yields a variety of patterns, including
arangements of asters, vortices and disordered
configurations.  We model this system {\it via} a
two-dimensional vector field describing the local
coarse-grained microtubule orientation and two
scalar density fields associated to molecular
motors. These scalar fields describe motors
which either attach to and move along
microtubules or diffuse freely within the
solvent. Transitions between single aster, spiral
and vortex states are obtained as a consequence
of confinement, as parameters in our model are
varied. We also obtain other novel states,
including ``outward asters'', 
distorted vortices and lattices of asters on
confined systems.  We
calculate the steady state distribution of bound
and free motors in aster and vortex
configurations of microtubules and compare these
to our simulation results, providing qualitative
arguments for the stability of different patterns
in various regimes of parameter space.  We also
study the role of crowding or ``saturation''
effects on the density profiles of motors in
asters, discussing the role of such effects in
stabilizing single asters. 
\end{abstract}
\date{\today}
\pagebreak
\maketitle

\section{Introduction}
The mitotic spindle in a dividing eukaryotic cell
is comprised of several millions of interacting
protein molecules\cite{alberts}. Remarkably,
these molecular constituents self-organize to
yield patterns at the scale of microns.  The
existence of such self-organized non-equilibrium
structures at the sub-cellular scale is a common
feature of biological systems.  Such structures
include the endoplasmic reticulum and the
Golgi complex, membrane-bound organelles which
participate in intra-cellular trafficking.
They also include the cytoskeleton, a
cell-spanning network of polymers such as
microtubules, actin filaments and intermediate
filaments\cite{alberts}.

he mitotic spindle consists of two
interpenetrating arrays of microtubules\cite{karsenti}. Each
such array originates from a microtubule
organizing centre formed at the two ends of the
dividing cell.  Microtubules are semi-flexible
polymers formed by polymerizing an asymmetric
dimeric unit\cite{dynins}. An individual
microtubule is a polar object: microtubule ends,
labelled as $-$ and $+$, grow and shrink at
different rates.  This polarity dictates the
direction of motion of a class of molecular motor
proteins on microtubules.  Motor proteins such as
kinesins use energy derived from adenosine
tri-phosphate (ATP) hydrolysis to exert forces
and to translocate along microtubules\cite{motor}. The
direction of motion is dictated by the filament
polarity -- both plus-end and minus-end directed
motors exist\cite{alberts,motor}.

The principal microtubule organizing centre in
most animal cells is the centrosome.  In a
dividing cell, the centrosome first duplicates to
form the two poles of a mitotic spindle. Each
centrosome nucleates a radial array of
microtubules called an {\em aster}. The two
asters move apart to opposite ends of the cell as
mitosis proceeds. Microtubules in the region
between them then preferentially elongate,
forming the spindle\cite{karsentivernos}.  

Interestingly, experiments on centrosome-free
fragments of the cytosol containing both motors
and microtubules also obtain self-organized
aster-like structures\cite{rodionovborisy}.  Asters, in
addition to other complex patterns such as vortices
and disordered aster-vortex mixtures, are also seen
{\it in vitro}, in experiments on mixtures of
molecular motors and
microtubules\cite{nedelecsurreymaggsleibler}.
Features of spindle formation are
reproduced in mixtures of motors, microtubules
and gold beads coated with DNA\cite{heald}.  The
fact that such experiments are
able to mimic the complex self-organized states
seen in living cells indicates
that simple mesoscale models which work with
fewer components may be useful in capturing some
aspects of cellular pattern formation
\cite{hopfield}.

This paper presents a theory of pattern
formation in mixtures of molecular motors and
microtubules in a confined geometry.  We
motivate hydrodynamic equations of motion for a
coarse-grained field representing the local
orientation of microtubules as well as for local
motor density fields\cite{physicascripta}.
In our theoretical description, as well as in the
experiments, microtubules are oriented by
complexes of bound motors, yielding patterns at
large scales\cite{elsewhere,hydrodynamic}.

The experimental work of relevance to this paper
is the following:  N\'ed\'elec and collaborators
study pattern formation in mixtures of complexes
of conventional kinesins with microtubules
in a confined quasi-two-dimensional geometry
\cite{nedelecsurreymaggsleibler}.
The later experiments of Surrey {\it et al.} investigate
pattern formation in larger systems where the
effects of boundaries appear negligible
\cite{surreynedelecleiblerkarsenti}.
The experiments are supplemented by
simulations which reproduce many
features of the experiments
\cite{nedelecsurreymaggsleibler,surreynedelecleiblerkarsenti,
nedelecsurrey,nedelecjcb,nedelecsurreykarsenti}.

Our approach is closest in spirit to that of Lee
and Kardar(LK) \cite{leekardar}.  The LK model
is a hydrodynamic description of two coupled
fields. One of these is a two-dimensional
vector field describing the coarse-grained
orientation of the microtubule. The other is
a conserved scalar field representing the
local motor density. The LK model captures
two prominent features of the experiments
of N\'ed\'elec {\it et al.}: the presence of
stable vortices formed by microtubules and the
instability of an aster formed in the early
stages of pattern formation to a single, stable
vortex at large motor densities.  However,
despite this success of the LK model, several
features of the experiments are incompletely
understood. The transition between a single
aster and a single vortex seen in the experiments
appears to be driven primarily by confinement.
(When a growing microtubule hits the wall of
the confining region, it bends and distorts.
The elastic energy cost of this distortion can
be relaxed in part if the microtubules assume a
vortex configuration.) In the LK model, confinement
does not appear to play a vital role, and the
single aster to single vortex transformation is
a generic feature even for large system sizes,
provided the motor density is sufficiently large.

Experiments obtain a variety of stable steady
states on larger systems as a function of motor
density.  These include a ``lattice of asters''
state in which asters are the only stable
structures, a ``lattice of vortices'' state,
in which individual vortices are stable while
asters are absent, as well as an intermediate
``aster-vortex mixture'' state.  The LK approach
predicts that a single vortex should be the
stable state at large motor densities even for
very large systems.  Experiments, however, always
see a ``lattice of asters'' in this regime.

The LK model predicts that motor density profiles
in asters are always simple decaying
exponentials, independent of the rates at which
motors hop on and off the filament.  However,
experiments and theoretical work suggest more
complex decays. Such decays include the
intriguing possibility of power-laws with
exponents which vary continuously as a function
of the on-off hopping rates.  N\'ed\'elec, Surrey
and Maggs (NSM) derive equations for motor
profiles around a single preformed aster, showing
analytically that such profiles are pure
power-law in nature\cite{nedelecsurreymaggs}.
The NSM equations describe motors as either
freely diffusing in the solvent or as bound to
and moving along a microtubule.  These states are
allowed to interconvert.  The densities of bound
and free motors are governed by separate
equations.  LK, in contrast,
use a single equation for the full motor
density field which is effectively the ``sum'' of
these equations but ignore the dynamics of the
difference field.

The NSM equations describe a {\em single} aster
configuration composed of a fixed number of
inward (or outward) pointing microtubules.  As a
consequence, the density of microtubule {\em
decreases} radially outward from the aster core.
Since the free motors become bound only in the
presence of microtubule, the conversion rate
should be proportional to the local microtubule
density.  This yields a non-trivial space
dependence for this conversion rate, which is
responsible for the power-law decay of motor
densities.  However, NSM do not address issues of
pattern formation.  We suggest here, in contrast
to NSM, that at the length scales appropriate to
a coarse-grained hydrodynamic description of
pattern formation and at the densities of
microtubules in the experiments, it is
appropriate to ignore fluctuations in the local
density of microtubule.  The relevant fluctuating
field is then the {\em orientation} field for the
microtubule at fixed density, as in the LK
model\cite{leekim,critique}.

Detailed information regarding the
ordering of microtubules in the presence
of molecular motors has come from the
extensive simulations of N\'ed\'elec and
collaborators \cite{nedelecsurreymaggsleibler,
surreynedelecleiblerkarsenti,nedelecsurrey,nedelecjcb}.
These simulations, performed in a two-dimensional
geometry, obtain asters and vortices, in addition
to relatively disordered configurations in which
both asters and vortices are present.  The best
simulations require as many as 19 parameters to
be specified; these include fluid viscosities,
motor diffusion constants, binding strengths
and microtubule bending rigidities.  However,
the uncertainties in these parameters are large,
often of an order of magnitude or more.

The hydrodynamic approach used here involves far fewer
free parameters\cite{kruse}. Since structures
obtained in the simulations are appropriate to
specific parameter values simulated, it is hard
{\it a priori}, to guess at what patterns would
result on changing one or more of them.  However,
one might expect several microscopic interactions
to yield the same hydrodynamic description,
provided it is general enough\cite{kruse}.
Simulating hydrodynamic equations is often
far faster and more efficient computationally
than direct molecular dynamics simulations. This
makes efficient scans of parameter space possible
\cite{tonertu}. Finally, simpler descriptions in
terms of a small set of model equations enable
analytic progress in some limits.

We summarize our results here: In a regime of
parameter space for our model which is closest
to that for the LK model, we obtain a single
vortex as a stable final state for large motor
densities.  Our results here coincide with
the LK results.  However, in other regimes,
asters are favoured.  A ``lattice of asters''
state is stabililized in our model through
a low-order relevant term in the equation
of motion for the microtubule orientation.
On small systems, constraints due to confinement
favour a small number of asters, whose number
can be increased systematically as parameters
are varied.  We provide qualitative arguments
for the stability of different patterns in varied
regimes of parameter space.

We calculate the distribution of free and bound
motors in asters and vortices obtained in our
model.  Our results for motor profiles about
asters differs from both the LK result and the
NSM one.  We derive an exponential decay of bound
motor densities away from aster cores, modulated
by a power-law in which the exponent of the power
law depends in a non-universal way on dynamical
parameters.  The associated decay length for the
exponential can become very large in some regimes
of parameter space, yielding what would appear to
be pure power-law decays close to 
the aster core.  In contrast, LK
would predict a pure exponential decay, whereas
the NSM result is a pure power law. For vortices,
we obtain results equivalent to the LK results.

We obtain, numerically, the solutions  to our
equations when ``crowding'' effects due to the
interactions of motors moving on microtubules
are accounted for in a simple way.  Such effects
distribute the motor density more uniformly along
the microtubules.  We argue that the inclusion of
such ``crowding effects'' should further act to
favour asters over vortices in finite systems. We
suggest that the term ``spiral'' as opposed
to ``aster'' or ``vortex'' is a more generic
description of the microtubule configuration.
We adduce simulational evidence for such spiral
structures favoured by confinement and point
out that the microtubule configurations seen in
experiments do resemble spirals in many cases.

\begin{figure*}
\begin{center}
\includegraphics[scale=0.75]{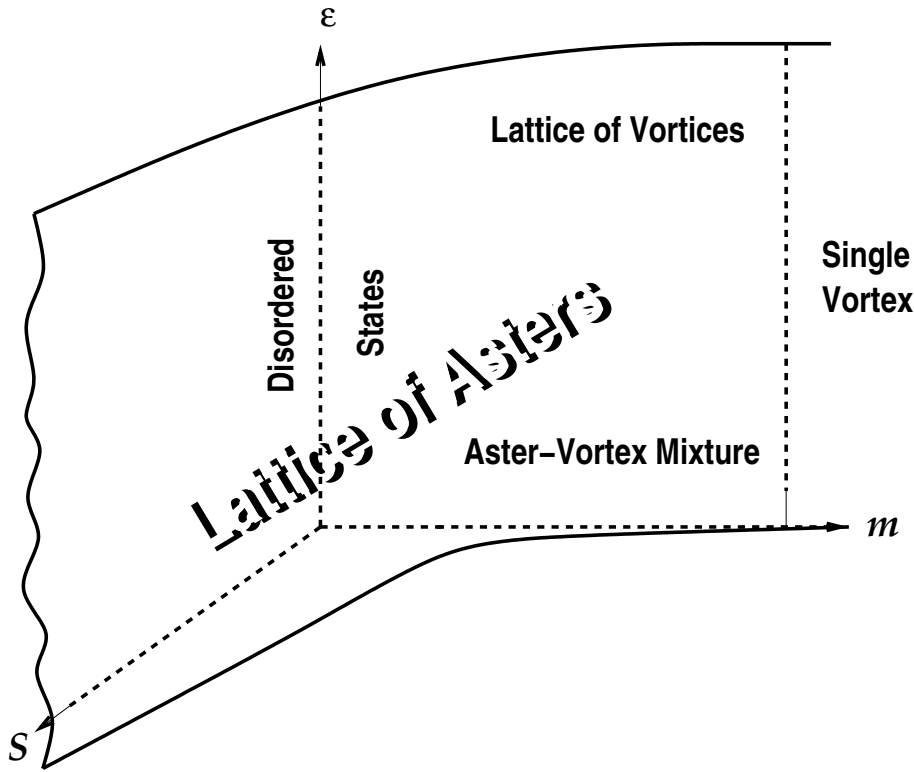}
\end{center}
\caption{\label{phasedia}
Qualitative ``phase diagram'' illustrating how
different states, the disordered state, the
aster-vortex mixture state, the lattice of
vortices state, the single vortex, the lattice of asters,
dominate in different regimes of parameter space;
for a definition of parameters see text. The
parameter $\epsilon$ is plotted on the $y$ axis,
with the total motor density $m$ plotted on the
$x$ axis. The parameter $S$ extends out of the
plane of the figure.}
\end{figure*}

Our results relating to pattern formation in
systems where the effects  of boundaries can be
neglected appear elsewhere. To summarize these
briefly, we show in Ref.\cite{elsewhere} that
our model generates all the patterns seen in
experiments, such as the aster-vortex mixture,
the ``lattice of asters'' and the ``lattice
of vortices''\cite{fail}.  We show how these
states are linked in a non-equilibrium ``phase
diagram'', reproduced in Fig. 1, demonstrating
how smooth trajectories in parameter space can
connect the states observed, in agreeement with
the experiments\cite{notephase}.  Fig. 1 is
relevant to experimental data in that it shows
how a relatively small number of parameters
may suffice to fix the macroscopic state of
the system.

The outline of this paper is the following: In
Section II, we describe the details of our model.
We discuss and justify each term which appears in
our equations of motion for microtubule and motor
fields. Section III describes our numerical
methods.  Section IV presents results from our
numerical simulation of the equations of motion
for different boundary conditions.  In Section V,
we present analytic results for the profile of
motor densities in single vortex and aster
configurations. In a subsection, these analytic
results are compared with results from direct
simulations of the hydrodynamic equations.
Section VI discusses how saturation effects
resulting from the interactions of bound motors
affects motor density profiles. Section VII
combines the results of Sections V and VI in a
discussion of the relative stability of aster and
vortex configurations. We provide a simple intuitive
argument for the stability of different patterns
as parameters in our model are changed. The concluding
section, Section VIII, summarizes the results of this
study and outlines possibilities for further
work.

\section{Model}

Our model treats motors  attached to microtubules
differently from motors which diffuse freely in
solution.  Motors which move on microtubules are referred
to as  ``bound''  motors, while those which diffuse in the
ambient solvent are referred to as ``free'' motors. These
are described by coarse-grained fields denoted by $m_b$ and
$m_f$ respectively and obey different  equations of
motion.  In the absence of interconversion terms changing a
bound motor to a free motor, $m_b$  obeys a continuity
equation involving the  current of motors transported along
the microtubules. The free motor field $m_f$ obeys a
diffusion equation with a diffusion constant $D$.  These 
two fields are  coupled  through
mechanisms which convert ``free''  motors to ``bound''
motors and vice versa. The equations they obey have the
form
\bea 
\partial_t m_f = D \nabla^2 m_f - \ftob^{\prime} m_f + \btof^{\prime} m_b \\
\partial_t m_b = -A \nabla.(m_b {\bf T}) + \ftob^{\prime} m_f -\btof^{\prime} m_b
\eea
$\ftob^{\prime}$ and $\btof^{\prime}$ are the rates at which free motors
become bound motors (``on'' rate) and vice-versa (`off'' rate).
These rates have dimensions of inverse time. Note that the {\em total}
motor density field $m = m_b(r) + m_f(r)$ is a conserved
field.

The term $A \nabla \cdot (m_b {\bf T})$ describes the motion of bound
motors along microtubules with velocity $A$. One could, as we
will do later, make the velocity itself density dependent.
(This is natural when modeling motor motion in
terms of the one-dimensional driven diffusive
dynamics of interacting particles
for which the current $J(\rho)$ is a non-linear
function of the density $\rho$.)
While the rate $\ftob^{\prime}$ should depend on the local
density of microtubule, we will assume that microtubule
density fluctuations are suppressed at the scales
relevant to a hydrodynamic description and
retain only the orientational degree of freedom of the
local microtubule field. We work in two dimensions throughout,
since the experiments were performed in a quasi-two dimensional
geometry\cite{hydrodynamic}.

The dynamics of the microtubules, given by the
equation below, incorporates the terms used by
Lee and Kardar in specific limits.
It also includes one completely new
term. As in the LK
model, we ignore fluctuations in the density of
microtubules, concentrating on their
orientational degrees of freedom.  The
hydrodynamic equation includes terms which
reflect the dynamics of individual microtubules.
We take these to be stabilized at unit length.
It also includes motor-independent and
motor-dependent orientation terms.  In principle,
for a non-equilibrium problem, all symmetry
allowed terms must figure here. Of these terms,
we will incorporate only the lowest-order symmetry 
allowed terms whose contributions can be justified 
transparently on physical grounds.

Our equation then reads,
\bea
\partial_t {\bf T} = {\bf T}(\alpha - \beta T^2) + \gamma m_b \nabla^2 
{\bf T} +  \gamma^{\prime} \nabla m_b \cdot \nabla {\bf T}  
+ \kappa^{\prime} \nabla^2 {\bf T}+
S^{\prime} \nabla m_b
\eea
The first term, ${\bf T}(\alpha - \beta T^2)$,
governs the stabilization  of the
microtubules at a preferred length of $\alpha/\beta$, which
we will normalize to unity. The  second and third terms,
$\gamma m_b \nabla^2 
{\bf T} +  \gamma^{\prime} \nabla m_b \cdot  \nabla {\bf T}$, 
are alignment terms,
reflecting the alignment of microtubules due to the action
of bound motors. The third term is also interpretable
as a ``convective'' term, in which the local velocity
which convects fluctuations in the {\bf T} field is
proportional to the gradient of bound motor density.

The fourth term, $\kappa^{\prime} \nabla^2 {\bf T}$,
describes an intrinsic
stiffness against distortions, allowing tubules to form an
ordered phase in the absence of thermal noise.  (This term
was dropped by LK, where it was assumed that the
orientation of the tubules was solely driven by the motor
current. We include it here because it is certainly
symmetry allowed, is linear in the field $T$ and thus
should appear at lowest order. It is also useful in maintaining
the local smoothness of the patterns generated). 

The last term is completely new; it is a symmetry allowed term of
linear order in the fields. Note that a few
terms allowed by symmetry have been intentionally
excluded. These include the term $m_b \nabla (\nabla \cdot T)$; 
such a term has an effect equivalent to the $\nabla m_b$ term 
of the equation above which is lower order in gradients.

In the LK model the coefficients $\gamma$ and
$\gamma^{\prime}$ are taken to be equal. Thus,
the second and third terms can be interpreted in
terms of the functional derivative of a ``free
energy'' term. In general however, away
from thermal equilibrium, these two
coefficients differ and we may explore
the regimes of parameter space in which their
relative strengths vary.  For convenience we
choose $\gamma^{\prime} = \epsilon \gamma$ and
vary the parameter $\epsilon$ to tune the
ratio of these two terms.

The following transformation simplifies 
the equations considerably:
scale length in units of $D/A \sqrt{\beta/\alpha} $, time in
units of $\beta D/(\alpha A^2)$, motor density in units of $D/\gamma$
and the tubule density in units of $\sqrt{\alpha/\beta}$.  The equations
then reduce to
\be
\label{scalefden}
\partial_t m_f = \nabla^2 m_f - \ftob m_f + \btof m_b 
\ee
\be
\label{scalebden}
\partial_t m_b = -\nabla.(m_b {\bf T}) + \ftob m_f - \btof m_b 
\ee
\be
\label{scalet}
\partial_t {\bf T} = C {\bf T} (1 - T^2) + m_b \nabla^2 {\bf T} 
+ \epsilon \nabla m_b \cdot \nabla {\bf T}  
+ \kappa \nabla^2 {\bf T}+S \nabla m_b 
\ee
The parameter $C$ given by $\beta D /A^2$ is the growth constant.
$\ftob$ and $\btof$ are scaled in units of inverse time (i.e. $(\beta
D/(\alpha A^2))^{-1}$. $\kappa^{\prime}$ is appropriately scaled to
$\kappa = \kappa^{\prime}/D$ and $S=S^{\prime} (\beta D)/(\alpha \gamma A)$.
Note that the scaled equations for the free and
bound motors (Eqns. \ref{scalefden},\ref{scalebden}) are invariant
when the motor densities are multiplied by a constant. We will use this
invariance in Section V(C) to compare analytic results for motor density 
profiles with results from numerical simulations.

We relate our scaled parameters to typical experimental
values in the following way: The tubule density, scaled in
terms of $\sqrt{\alpha/\beta}$, is chosen to be unity. The
diffusion constant $D$ is about $20 {\mu m}^2/s$ and $A
\sim 1 \mu m/s$, defining basic units of length and time
as $20 \mu m$ and $20$ seconds respectively. A tubule
density of $1$ implies that over a coarse-graining length
of $400 {\mu m}^2$, there are around $400$ microtubules, a
value close to that used in the simulations
\cite{surreynedelecleiblerkarsenti}.  
Our choice for $\ftob$ and $\btof$
corresponds to physical rates of
$0.005 s^{-1}$ to $0.05 s^{-1}$, slightly smaller than
those in the simulations\cite{nedelecjcb}; using 
larger rates does not affect our conclusions here.

\section{Numerical Methods}
We solve Equations\ \ref{scalefden},\ref{scalebden} and \ref{scalet} 
numerically on an $L \times L$
square grid indexed by $(i,j)$ with $i = 1,\ldots L$ and $j = 1,\ldots L$.
The equation for the free motor density is evolved through an
Euler scheme,
\begin{equation}
m_f(t + \Delta t) = m_f(t) - \Delta t \nabla \cdot J(m_f) 
- \ftob m_f + \btof m_b,
\end{equation}
where
\bea
J_x(m_f(i,j)) = (m_f(i+1,j) - m_f(i-1,j))/2\delta,  \\
J_y(m_f(i,j)) = (m_f(i,j+1) - m_f(i,j-1))/2\delta.
\eea
The grid spacing is $\delta x = \delta y = \delta = 1$ and the
time step $\Delta t = 0.1$.
At the boundaries, we impose the boundary condition that no current (either
of free or bound motors) flows into or out of the system. This
condition is easily imposed by setting the appropriate current to zero.

A related discretization is used for the bound motor density 
equation
\begin{equation}
\partial_t m_b = -\nabla \cdot (m_b {\bf T}) + \ftob m_f - \btof m_b,
\end{equation}
where, in the bulk, partial derivative terms are discretized as 
\begin{equation}
\partial_x (m_b {\bf T}) = (m_b(i+1,j)T_x(i+1,j)  
- m_b(i-1,j)T_x(i-1,j))/2\delta.
\end{equation}
with a similar equation used for the $y$-component.

The {\bf T} equation is differenced through the Alternate Direction 
Implicit (ADI) operator splitting method in the  Crank-Nicholson scheme.
At the first half time-step
\be
(r {\cal I} - {\cal L}_x ) T_x^{n + 1/2} = (r {\cal I} + {\cal L}_y) T_x^{n},
\ee
where $r = 2 \delta^2/\delta t$ and ${\cal I}$ is the identity matrix. The 
superscripts on $T_x$ indicate the time-step at which these quantities are 
calculated. The operators
${\cal L}_x$ and ${\cal L}_y$ are given by
\bea
{\cal L}_x = m_b \partial_x ^2 + (\partial_x m_b) \partial_x + 
S (\partial_x m_b),\\
{\cal L}_y = 2 C (1-T^2) + m_b \partial_y^2 + (\partial_y m_b) \partial_y+
S (\partial_y m_b). 
\eea
The first and second derivatives evaluated for a 
function $f(i,j)$ on a lattice point $(i,j)$ 
in the bulk are
\bea
\partial_x f(i,j) &=& (f(i+1,j)-f(i-1),j))/2\delta, \\
\partial_x^2 f(i,j) &=& (f(i+1,j)-2 f(i,j) + f(i-1,j))/\delta^2,
\eea
with similar equations for the $y$-derivatives.

At the second half time step
\be
(r {\cal I} - {\cal L}_y ) T_x^{n + 1} = (r {\cal I} + {\cal L}_x) T_x^{n+1/2},
\ee
where
\bea
{\cal L}_x = m_b \partial_x ^2 + (\partial_x m_b) \partial_x + 
S (\partial_x m_b), \\
{\cal L}_y =  m_b \partial_y^2 + (\partial_y m_b) \partial_y + 
S (\partial_y m_b).
\eea

A similar scheme is used for differencing the
equation for the $T_y$ component. Our simulations
are on lattices of several sizes, ranging from
$L=30$ to $L=200$. We vary the motor density in
the range 0.01 to 5 in appropriate dimensionless
units.  We work with two different types of
boundary conditions on the $T$ field. In the
first, which we refer to as reflecting boundary
conditions, the microtubule configuration at the
boundary sites is fixed to point along the inward
normal.  In the second, which we refer to as
parallel boundary conditions, microtubule
orientations at the boundary are taken to be
tangential to the boundary.  In both these sets
of boundary conditions, the state of the boundary
$T$ vectors is fixed and does not evolve. The
total number of motors, initially divided equally
between free and bound states and distributed
randomly among the sites, is explicitly conserved.

We add weak noise, primarily in the {\bf T}
equation of motion, to ensure that true steady
states are reached in our simulations. Large
noise strengths wipe out patterns, yielding
homogeneous states. We have also experimented
with a variety of initial states to ensure that
the qualitative features of the patterns we
obtain as stable steady states are, indeed,
robust.

\section{Results and Discussion: Numerical}

This section presents the results of our
simulations in different regimes of parameter
space for the boundary conditions discussed
above.  The simulations discussed here are on
systems of size $L = 30$, corresponding to
physical length scales of about $60\mu m$.
Our results on somewhat
larger systems ($L=50$) are intermediate in
character between the results for $L=30$
presented here and our results on systems of much
larger sizes ($L \ge 100$), where boundary
effects are negligible.

We work at a fixed large
motor density of $m = m_b + m_f = 0.5$. We work
at fixed values of $\btof
 = \ftob = 0.5$ here but have checked that making
the motors more or less processive does not alter
our results qualitatively. For motors with very
small ``on'' and ``off' rates, we see disordered
states best described as aster-vortex mixtures.
The patterns obtained here emerge at still higher
motor densities for low values of the
processivity. We work with a small value of
$\kappa$, ($\kappa = 0.05$ here),  such that the
magnitude of the self alignment term $\kappa
\nabla^2 {\bf T} $ is  small compared to $m_b
\nabla^2 {\bf T}$.

\subsection{Reflecting boundary conditions}

\begin{figure*}
\begin{center}
\includegraphics[scale=0.75]{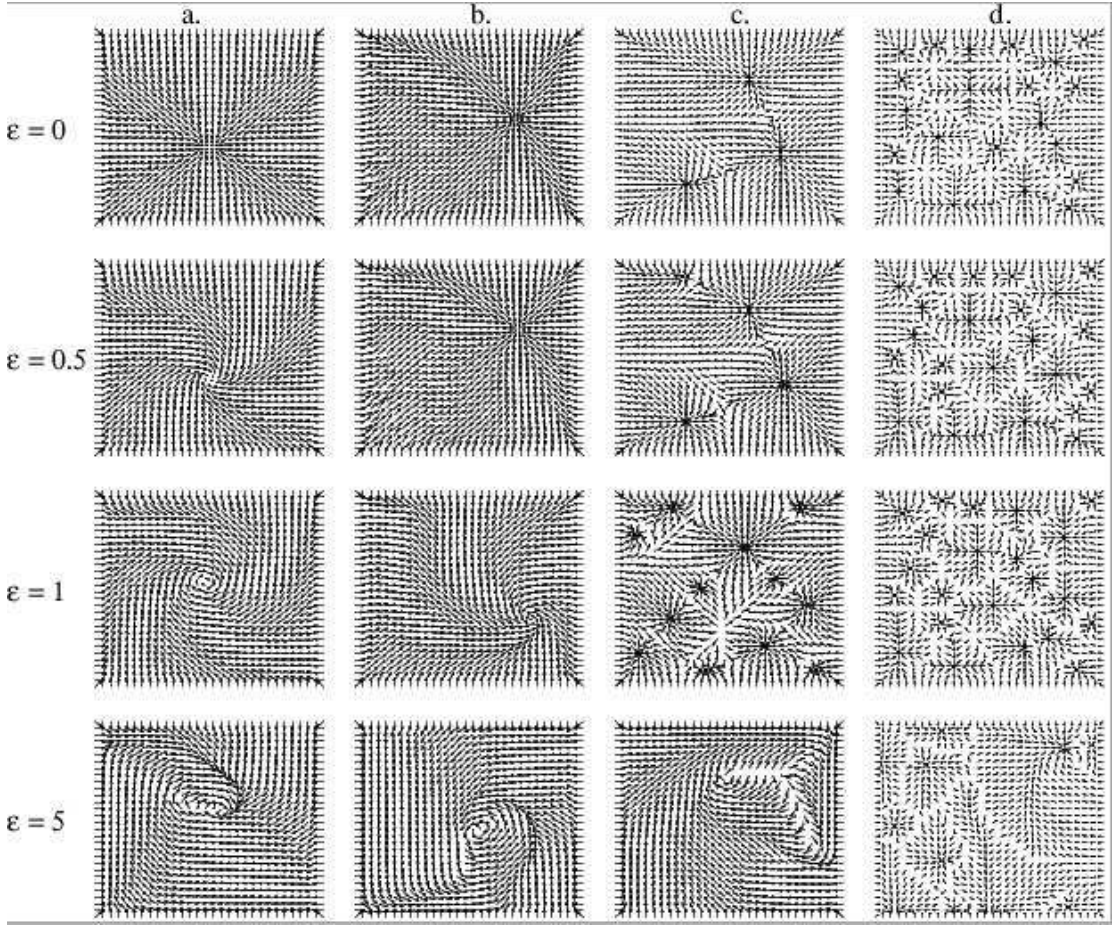}
\end{center}
\caption{\label{reftube}
Steady- state configurations of microtubules with reflecting 
boundary conditions.  The parameters are $m= 0.5$ and 
$\ftob=\btof=0.5$. Patterns are shown at different 
$\epsilon$ at (a) $S=0$, (b)$S=0.05$ ,(c)$S=0.5$ and
(d) $S=2$.}
\end{figure*}

Our results for reflecting boundary conditions
are shown in Fig. 2(a) -- (d). Each column
represents a different value of $S$ : {\it i.e.}
(a) $ S = 0$, (b) $S = 0.05$ (c) $S = 0.5$ and
(d) $S = 2$. We vary the parameter $\epsilon$ in
these scans, with $\epsilon$ taking the values
$\epsilon = 0.0, 0.5, 1.0$ and $5$ as shown.

In Figs. 2(a) -- (d), the effects of varying
$\epsilon$ with reflecting boundary conditions
are the following:  In Fig. 2(a), $S = 0$ and the
steady state configuration at $\epsilon = 0$ is
an aster with tubules directed radially inward
towards the core.  When $\epsilon$ is increased
to $0.5$, the configuration resembles a
``spiral''.  As $\epsilon$ is increased further
the spiral distorts into a vortex.  In Fig. 2(b),
we show configurations at small but non-zero $S$,
$S=0.05$.  The single aster is
stable for $\epsilon = 0$ and  $0.5$, but yields to
a single vortex for larger
$\epsilon$.  A general observation is that the
core of the vortex is increasingly distorted as
$\epsilon$ is increased further.

\begin{figure*}
\begin{center}
\includegraphics[scale=0.6]{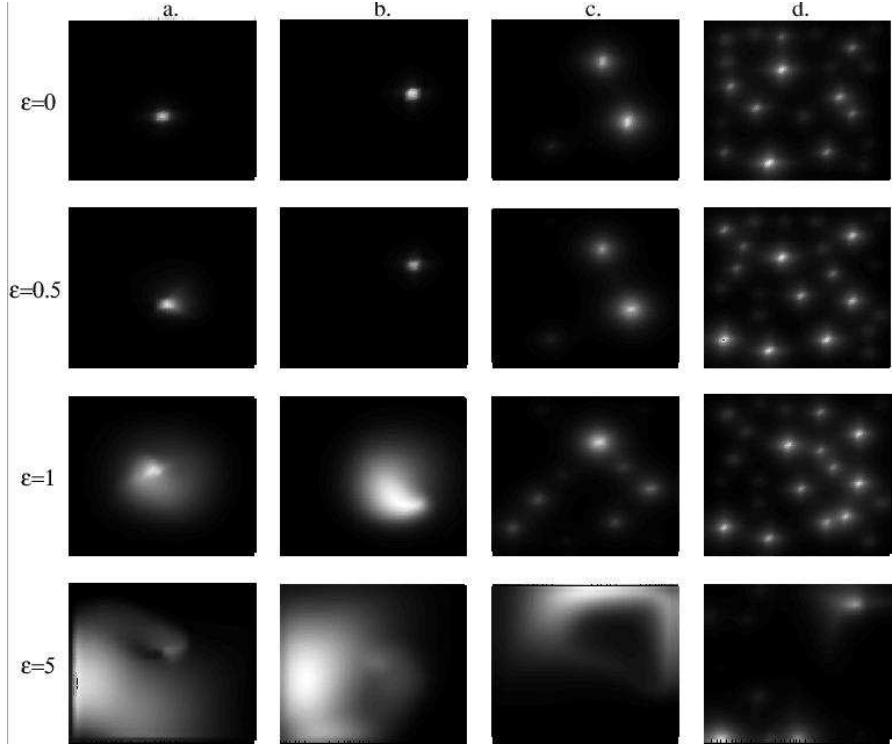}
\end{center}
\caption{\label{refdenb}
Steady- state bound motor density profiles with reflecting boundary conditions.
The parameters are $m= 0.5$ and $\ftob=\btof=0.5$. Profiles
are shown at different $\epsilon$ at (a) $S=0$, (b)$S=0.05$ ,(c)$S=0.5$ and
(d) $S=2$.The darker regions indicate regions of lower motor density.}
\end{figure*}

Fig. 2(c) shows steady state configurations at
$S=0.5$. A tendency towards the
formation of a lattice of asters is apparent,
as $\epsilon$ is
increased at these values of $S$. This is
consistent with our earlier results on large
systems, where we observed that non-zero $S$
always promoted the lattice of asters. At
large $\epsilon$ {\it e.g.} $\epsilon = 5$, there
is a pronounced tendency towards alignment,
yielding a steady state pattern of a vortex with
a highly distorted core.  Fig. 2 (d) shows
configurations at $S=2$, where the lattice of
asters is present at all values of
$\epsilon$ shown. At large $\epsilon$, the
tendency towards parallel alignment competes with
the tendency towards aster formation. At still
larger values of $\epsilon$, we obtain an aligned
phase in the bulk (not shown), reminiscent of the
``bundles'' seen in the experiments at large
motor concentration.

\begin{figure*}
\begin{center}
\includegraphics[scale=0.75]{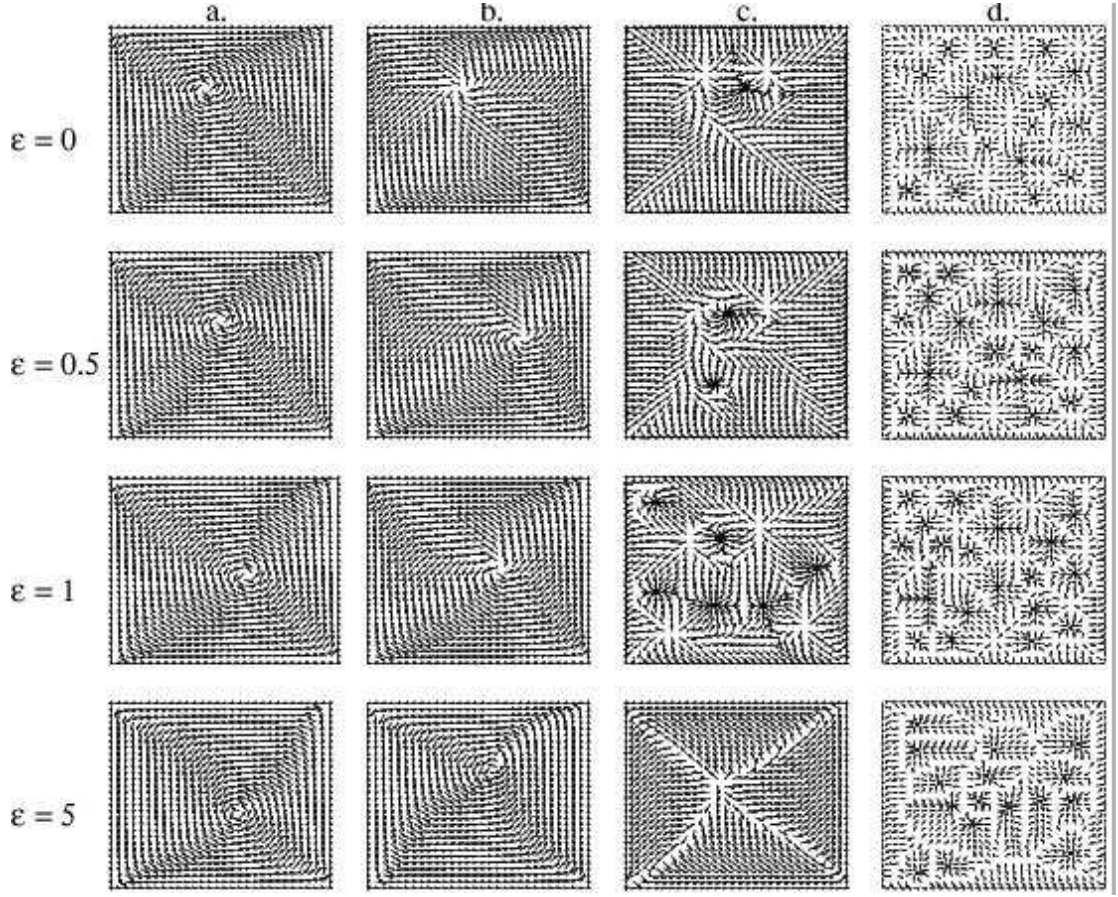}
\end{center}
\caption{\label{partube}
Steady- state configurations of microtubules with parallel boundary conditions.
The parameters are $m= 0.5$ and $\ftob=\btof=0.5$. Patterns
are shown at different $\epsilon$ at (a) $S=0$, (b)$S=0.05$ ,(c)$S=0.5$ and
(d) $S=2$.}
\end{figure*}

Figures 3(a) -- (d) show bound motor density
profiles corresponding to the microtubule
arrangements of Figs. 2(a) -- (d). Lighter
regions of the figure indicate regions of larger
density.  Note that the bound motor density is
concentrated at the centres of asters.  In vortex
configurations, the motor density profiles for
bound motors are far smoother. In the ``lattice
of asters'' configuration, the bound motor
density profiles peaks at the centres of the
asters, decaying to a small value at intermediate
points far away from aster cores.  The profiles of
free motor densities are visually very similar to
those for bound motors and are not shown here.

\subsection{Parallel boundary conditions}

\begin{figure*}
\begin{center}
\includegraphics[scale=0.6]{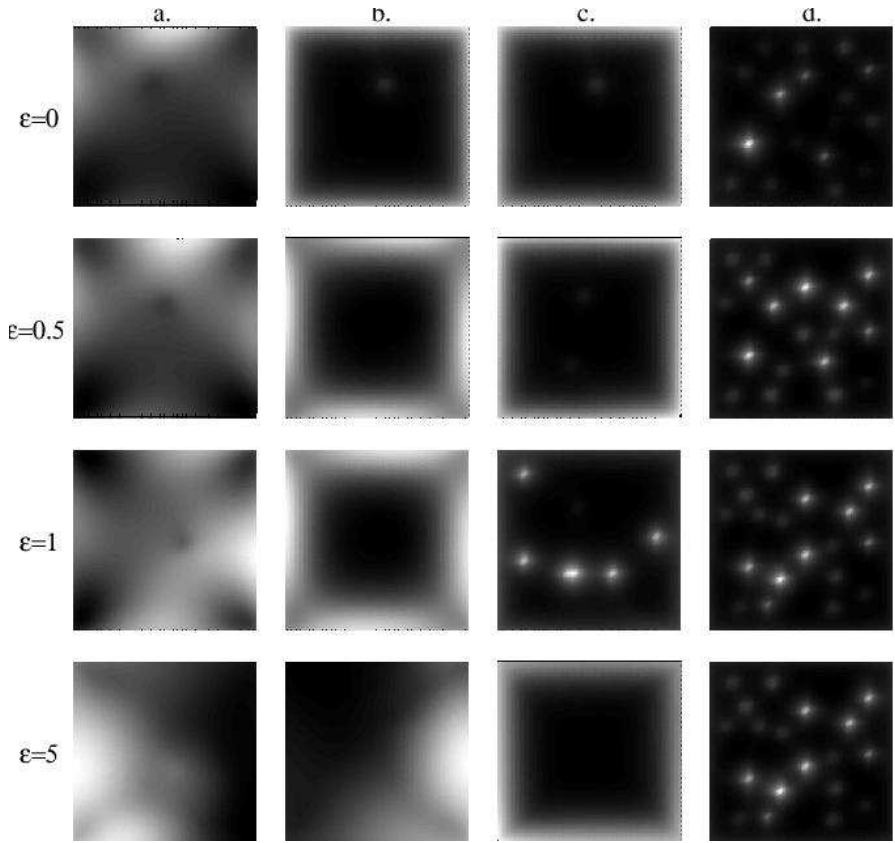}
\end{center}
\caption{\label{pardenb}
Steady- state bound motor density profiles with parallel boundary conditions.
The parameters are $m= 0.5$ and $\ftob=\btof=0.5$. Profiles
are shown at different $\epsilon$ at (a) $S=0$, (b)$S=0.05$ ,(c)$S=0.5$ and
(d) $S=2$.The darker regions indicate regions of lower motor density.}
\end{figure*}

We now discuss pattern formation in finite
systems with parallel boundary conditions, see
Figs. 4(a)--(d).  These are to be contrasted to
Figs. 2(a) --(d).  Note that qualitatively
different sequences of patterns are stabilized at
identical values of other parameters, depending
on the boundary conditions. This illustrates the
sensitivity to boundary conditions which obtains
for sufficiently small system sizes $L$.

Fig. 4(a) shows pattern formation with parallel
boundary conditions at $S=0$. A single vortex is
obtained as a steady state at all values of
$\epsilon$ shown {\it i.e.} $\epsilon =
0.0,0.5,1.0$ and $5$. Fig. 4(b) shows  patterns
for $S=0.05$.  The steady state here is a single
aster with tubules pointing {\em outwards} at the
core, but aligned with the configurations at the
boundary. (We term such configurations as
``outward asters''.) Clearly these arise as a
consequence of the boundary conditions which favour
vortices and the parameter regime which favours 
asters. At $\epsilon=5$, the steady
state is a well-formed, clean vortex. Fig 4(c)
illustrates pattern formation at $S=0.5$, as a
function of $\epsilon$. We see a tendency towards
the formation of the lattice of asters phase
expected for large $S$ for sufficiently large
$\epsilon$, although indications of this are seen
even for small $\epsilon$.

At very high $\epsilon$, we obtain an unusual
configuration which we term a ``flag''. In this
configuration, tubules point radially outward
near the core while merging with the imposed
parallel microtubule configuration near the
boundary. However, the configuration appears to
have a four-fold axis, in which the axis lines
are along the diagonal. Across this axis, the
microtubule orientation changes sharply. This
4-fold symmetry reflects the four-fold
symmetry of the simulation box.
Finally, Fig. 4(d) shows patterns formed at $S=2$.
In this regime, the lattice of asters is
the stable steady state for all values of
$\epsilon$.

Figures 5(a) -- (d) shows bound motor density
profiles corresponding to the microtubule arrangements
of Figs 4(a) -- (d). Note that the density distribution
in single vortices varies smoothly, consistent
with theoretical expectations. The profiles appear
sensitive to the boundary and a four-fold rotation
axis can be seen in several of the patterns which
involve a single vortex. The patterns for non-zero and
large $S$ are very similar to those obtained for
reflecting boundary conditions with similar values of
parameters.  The profiles of free motor densities are 
again visually very similar to those for bound motors 
and are not shown here.

\subsection{Discussion}

Our results for the cases outlined in the
previous sub-section are summarized as follows.
For reflecting boundary conditions, we obtain the
general sequence aster $\rightarrow$ spiral
$\rightarrow$ vortex at $S=0$.  This sequence is
obtained at fixed motor density, as a function of
the non-equilibrium parameter $\epsilon$. With
parallel boundary conditions, the patterns formed
are generically vortices, although the region
surrounding the core is progressively distorted
as $\epsilon$ is increased. Finite $S$ favours
the formation of a lattice of asters, as in large
systems. We observe some novel configurations such as the
``flag'' configuration, reflecting the four-fold symmetry
of the simulation box and the outward aster.

We commented earlier that the configurations
obtained in experiments appear to represent states
intermediate between asters and vortices in the
most general case. We expand on this here: 
A spiral configuration is described by
the unit vector field ${\bf T} = T_r{\hat r} 
+ T_\theta{\hat \theta}$ 
in which $T_r$ and $T_\theta$ take the forms,
\be
T_r=\cos(\alpha) \,\,,\,\, T_{\theta}=\sin(\alpha),
\ee
where $\alpha$ is a constant. This contains both asters and 
vortices in appropriate limits:  In the limit $\alpha=0$, this
equation describes  
an aster, while in the limit $\alpha=\pi/2$, the
configuration is a vortex. Thus asters and
vortices are particular limiting cases of more 
general spiral states.

In our simulations, spirals arise as a
consequence of the competition between the vortex
configurations favoured locally for intermediate
and large values of $\epsilon$ and reflecting
boundary conditions, which favour asters. Spirals
offer the best compromise between these and it is
reasonable to expect that such states should be
more generically seen than either asters or vortex 
states. Indeed, inspection of the configurations
associated to the ``lattice of vortices'' 
in Ref. \cite {surreynedelecleiblerkarsenti} provides
strong visual evidence for generic spiral states.

The bound and free motor profiles we obtain in
the simulations are consistent with expectations
from our theoretical analysis (see below). The
characteristic sharp peaks in motor density
profiles associated with asters are replaced by
far more slowly varying profiles for vortices.
The lattice of asters, therefore has strong
signals in the motor distribution function.
Appropriate experimental labelling of motors in
this phase should yield patterns and profiles
closely similar to those displayed here.

\section{Analytic Results for Motor profiles in
Aster and Vortex Configurations}

We now present an analytic calculation of free
and bound motor density profiles in vortex and
aster configurations.  We fix single vortex and
aster configurations of microtubules and solve
for steady state free and bound motor densities
in the presence of the fixed microtubule
configuration.

We find steady state solutions of Eqns.\ \ref{scalefden} and
\ref{scalebden} 
by setting the time derivatives to zero
{\it i.e.}\/ $\partial_t m_f = \partial_t m_b = 0$.  Adding and
subtracting these equations, we obtain
\be
\label{eq1} \nabla^2 m_f - \nabla \cdot (m_b {\bf T}) = 0,
\ee
\be
\label{eq2} \nabla^2 m_f + (\btof m_b - \ftob m_f) = 0.
\ee
The first of these equations implies that
\be 
{\bf \nabla} m_f = m_b {\bf T} + {\bf C}, 
\ee
where ${\bf C}$
is an arbitrary vector whose divergence ${\bf \nabla} \cdot {\bf
C} = 0$.

\subsection{Single vortex i.e. ${\bf T}$ = $\hat \theta$}
Assuming radial symmetry, we may write
$m_f = m_f(r)\ \textnormal{and}\ m_b = m_b(r)$.
Let ${\bf C} = c_r {\bf \hat r} + c_{\theta} {\bf \hat \theta}$. 
Eqn.\ \ref{eq1} then implies
\bea
\partial_r m_f \hat{r} = m_b \hat{\theta} + c_r \hat{r} 
+ c_{\theta} \hat{\theta}.
\eea
Hence,
\be
c_r = \partial_r m_f\  \textnormal{and}\  c_{\theta} = - m_b.
\ee
The constraint of zero divergence for ${\bf C}$ yields
\bea
\label{eqn3}
\partial_r c_r + \frac{c_r}{r} + \frac{1}{r} \partial_{\theta} c_{\theta} = 0.
\eea 
Since $c_{\theta} = -m_b(r)$ is purely radial
\bea
\partial_{\theta} c_{\theta} = 0.
\eea
Eqn.\ \ref{eqn3} then becomes a radial equation
\bea
\partial_r c_r + \frac{c_r}{r} = 0.
\eea
Using $c_r = \partial_r m_f$, we obtain
\bea
{\bf \nabla}^2 m_f = 0,
\eea
which supports solutions of the form
\bea
m_f(r) = c_1 + c_2 \ln(r),
\eea 
where $c_1$ and $c_2$ are constants to be determined by 
boundary conditions and normalization. 

Eqn.\ \ref{eq2}  now simplifies to
\be
\btof m_b - \ftob m_f = 0  \nn
\ee
which yields
\be
m_b(r) = \frac{\ftob}{\btof} (c_1 + c_2 \ln(r))
\ee
We must determine the constants $c_1$ and $c_2$. There
is no radial current of motors (bound or free) in the 
vortex configuration.
The condition of no radial current of free
motors at the boundary implies
\be
\partial_r m_f(r) = 0\ \textnormal{at}\ r = L,
\ee
therefore implying that $c_2 = 0$. The constant $c_1$ is determined 
by normalization
\be
\int d^2 r (m_f(r)+m_b(r)) = N.
\ee 
In the vortex configuration
\be
\int d^2r (1+\frac{\ftob}{\btof}) c_1 = N,
\ee
which gives
\bea
c_1 = \frac{n}{2 \pi} \frac{1}{(1+\frac{\ftob}{\btof})},
\eea
with $n$ the density of motor $N/L^2$.

The result above holds for purely tangential
boundary conditions, consistent with the symmetry
of the vortex. It will be modified {\it vis. a
vis} the simulation results both as a consequence
of the (cartesian) latticization used to
discretize the equations as well as by the
four-fold symmetry of the simulation box.  Both
these factors will lead to a combination of the
logarithmic and constant solutions above.

\subsection{Single aster solution i.e. ${\bf T} = -{\bf \hat r}$}
We again assume radial symmetry for the bound and free motor densities. 
We choose ${\bf C} = f(r) \hat r$, with $f(r)$ 
such that ${\bf C}$ is
divergenceless {\it i. e.} $\nabla \cdot {\bf C} = 0$. 
We thus obtain $f(r) = c_0/r$, with $c_0$ a 
constant.
The radial current of free motors is $-\partial_r m_f \hat r$ 
while the radial current of bound motors is $-m_b \hat r$. The 
boundary condition that the total motor current
vanishes at the boundary implies
\be
\partial_r m_f + m_b = c_0/r = 0 ~~~at~ r=L.
\ee
This ensures that $c_0 = 0$ and therefore
\be
\label{astercondition}
\partial_r m_f(r)  = - m_b(r).
\ee
Eqn.\ \ref{eq2} becomes
\be
\label{asterfree}
\partial_r^2 m_f + (\frac{1}{r} - \btof) \partial_r m_f - \ftob m_f = 0.
\ee
This is a second order differential equation whose solution
is completely specified if the value of the function and of its 
derivative at the boundary are supplied. Given these boundary 
conditions, the free motor density can be obtained via a numerical 
solution using the Runge-Kutta method. 
The bound motor density is also easily determined {\it via}
a radial derivative of the free motor density.

We now derive exact and asymptotic expressions for motor
densities in the aster geometry.  The general solution to
the equation above is a combination of confluent hypergeometric
functions and has the form
\bea
\label{astersolutions}
m_f(r) = e^{\expo~r} [c_2 {}_1F_1(\fa,1, \ta)+ \nn \\ 
c_1 U(\fa,1,\ta)].  
\eea
The
functions ${}_1F_1(\fa,1,\ta)$ and $U(\fa,1,\ta)$ are the two solutions of the
confluent hypergeometric Kummer equation. It is useful to define a
quantity $p$ given by
\be
p = \frac{1}{2}(1 - \frac{\btof}{\sqrt{\btof^2+4\ftob}}).
\ee
Note that
\be
0 \le p \le 0.5,
\ee
with $\btof,\ftob \ge 0$. The argument of the exponent 
in Eq. \ \ref{astersolutions} is always
negative for all $\btof,\ftob
\ge 0$. The coefficients $c_1$ and $c_2$ are determined by
boundary and normalization conditions.

The boundary condition is that there is no total motor current 
at the boundary. i.e. 
\be 
\partial_r m_f(r) + m_b(r)  = 0\ \textnormal{at}\  r=L,
\ee
since the radial current of bound motors in the
aster configuration is ${\bf J} = - m_b(r) \hat{r}$.
This can be used to determine $c_2$ in terms of
$c_1$.
We find that on imposing the no-current boundary condition, $c_2$ is
very small compared to $c_1$ and is significant only when
the system size is of order the ``correlation'' length
over which the density decays. We will assume (see below) that
we can set $c_2 = 0$ to yield physically admissible solutions.

The constant $c_1$ can now be fixed by the
normalization condition which requires the total number of motors
(bound+free) be constant,
\be
\int^{L}_{0} d^2 r (m_b(r)+m_f(r)) = N.
\ee
where $N$ is the total number of motors.

A physical argument for neglecting the ${}_1F_1(\fa,1,\ta)$ term in
comparison with the $U(\fa,1,\ta)$ is the following.  A
power series expansion yields
\be
{}_1F_1(p,q,x) = \sum_{n=0}^{\infty} \frac{(p)_n}{(q)_n} \frac{x^n}{n !}.
\ee
where
$(p)_n = (p+n-1)!/(p-1)!$ and $(p)_0 = 1$.  This function is
a monotonically (exponentially) increasing function of $r$.  
Such a density distribution
implies a motor density per unit area which increases as $r$ is
made arbitrarily large, a solution which is clearly untenable
on physical grounds. On the other hand, 
$U(\fa,1,\ta)$ decreases monotonically with increasing $r$.
Hence we retain
only the $U(\fa,1,\ta)$ part of the solution for $m_f(r)$.  

To derive the asymptotics we begin with the integral 
representation
\be 
U(p,q,x) =
\frac{1}{\Gamma(p)} \int_{0}^{\infty} e^{-xt} t^{p-1} (1+t)^{q-p-1} dt.
\ee
We relabel $(\sqrt{\btof^2+4\ftob}) r = z$ and $\frac{1}{2}(1-\frac{\btof}
{\sqrt{\btof^2+4\ftob}}) =
p$ for notational convenience.
We then obtain
\bea
U(p,1,z) = \frac{1}{\Gamma(p)} \int_{0}^{\infty} 
\frac{e^{-zt}~t^{p-1}}{(1+t)^{p}} dt, 
\eea
We now change variables to $u = zt$. The integral is then
\be
\frac{1}{\Gamma(p)} \frac{1}{z^p} 
\int_0^{\infty} \frac{e^{-u} u^{p-1}}{(1+u/z)^p} du.
\ee
Since $p$ is a small number between $0$ and $0.5$ we may 
expand the denominator binomially.  The first term gives
$\Gamma(p)$ and subsequent terms converge. Hence in the large $z$
(asymptotic limit), the integral is just $\Gamma(p)$. Therefore,
\bea
m_f(r)  &=& c_1 \frac{e^{-r/\xi}}{{(\btof^2+4\ftob)}^{p/2} r^p} \nn \\
m_b(r) &=& c_1 \frac{e^{-r/\xi}}{{(\btof^2+4\ftob)}^{p/2} r^p}
(\frac{p}{r}+\frac{1}{\xi}), \nn
\eea
where the inverse of the ``correlation'' length is defined as
\be
\xi^{-1} = \Big|\frac{(\btof - \sqrt{(\btof^2+4\ftob})}{2}\Big| = 
\Big|\frac{p\btof}{2p-1}\Big|.  
\ee
The correlation length $\xi$ and the power-law exponent $p$
depend on $\ftob$ and $\btof$.
We see that the bound motor density in the aster case has an
exponential fall modulated by a power-law tail instead 
of the pure exponential falloff predicted in the LK model.

Another limit in which an exact answer can be obtained is
the following: If there is no interconversion
between the two species of motors ({\it i.e.} $\btof = \ftob = 0$),
the equations of motion for free motors and bound
motors decouple.  The number of free motors and
bound motors in the system are then conserved independently.
The free motor density equation supports solutions of the form $m_f
(r) = c_1 + c_2 \ln (r)$, where $c_1$ and $c_2$ are
constants of integration. Imposing a vanishing
current of free motors at the boundary constrains $c_2 =
0$. The other constant $c_1$ can be fixed from the
condition that the total number of free motors is
constant.  The bound motor density then has the
simple power-law behavior $m_b(r) \sim  1/r$.

\subsection{Comparison of the analytic results with simulations}

\begin{figure*}
\begin{center}
\includegraphics[scale = 0.5]{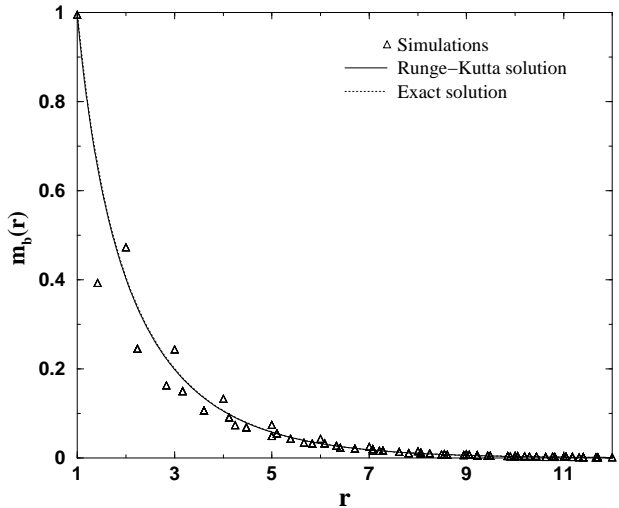}
\end{center}
\caption{\label{compb}
Density profiles for bound motor densities. The profiles 
are shown for an aster in the regime $\ftob = \btof = 0.5$,
with $S=0$.  The density is chosen to be $0.5$.}
\end{figure*}

These analytic results can now be compared 
to the results of direct simulations of the
dynamical equations for single aster and vortex
structures.  We choose ${\bf T} = \hat r$
in the full equations of motion for the motor
densities given in Eqns.\ (\ref{scalefden}) and
(\ref{scalebden}) to represent the aster. We then
evolve these equations in time till the steady
state is reached.  We compare the motor profiles
obtained in such simulations to profiles obtained
from solving the ODE at steady state
(Eq.\ \ref{asterfree}) by a 4$^{th}$ order
Runge-Kutta method.  Free and bound motor
densities at the boundary of the system are
obtained by solving the full time-dependent
equations
(Eqns.\ \ref{scalefden} and \ref{scalebden}). These
are used as boundary value input to the
Runge-Kutta procedure to facilitate
direct comparison between simulation and analytic
results.

\begin{figure*}
\begin{center}
\includegraphics[scale = 0.5]{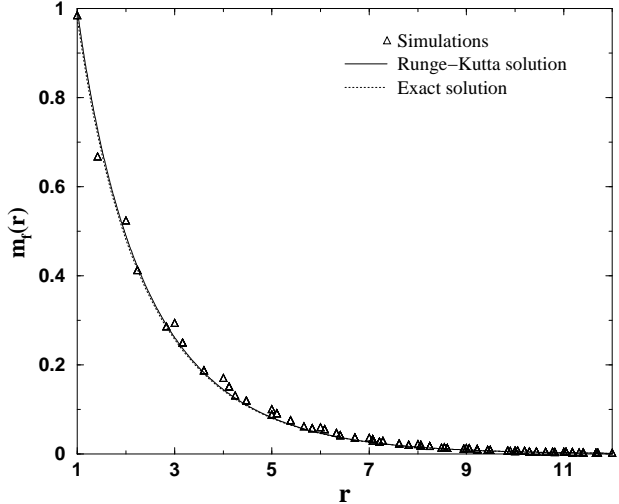}
\end{center}
\caption{\label{compf}
Density profiles for free motor
densities. The profiles are shown for an aster in the
regime $\ftob = \btof = 0.5$ and $S=0$.
The density is chosen to
be $0.5$.}
\end{figure*}

Using the symmetry of the scaled equations Eqns.\ \ref{scalefden} and
\ref{scalebden} discussed in Section II, we normalize the densities
obtained by these two methods to the same constant value before 
comparing them.  Figure 6 shows a plot of the distribution of bound motors
in a single aster obtained from (i) the direct simulation
(ii) the Runge-Kutta method and (iii) the exact solution
outlined in the previous section. The parameter values
are $\ftob = \btof = 0.5$ and $m=0.5$. Fig. 7 shows the
comparison between simulations, the Runge-Kutta method and
the exact solution for the free motor density profiles.
While the data is noisy, particularly for the bound motor
densities, the agreement with the theoretically predicted
result is satisfying.

\begin{figure*}
\begin{center}
\includegraphics[scale = 0.5]{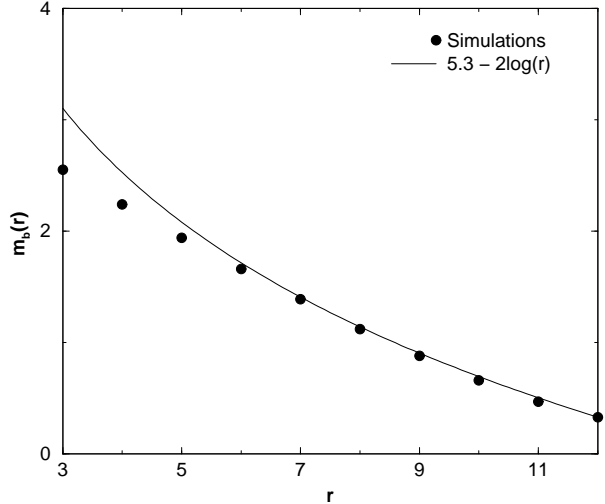}
\end{center}
\caption{\label{vortprof}
Density profiles for bound motor
densities. The profiles are shown for a vortex in the
regime $\ftob = \btof = 0.5$ and $S=0$.
The density is chosen to
be $0.5$.}
\end{figure*}

Fig. 8 shows the distribution of bound motors in
a vortex configuration obtained by fixing ${\bf
T} = \hat \theta$ in the full equations of motion
for the motor densities given in
Eqns.\ \ref{scalefden} and \ref{scalebden}. The plot,
of $m_b(r)$ {\it vs.} $r$, illustrates the slow,
essentially logarithmic variation of the motor
densities in such configurations, consistent with
the analytic approach of this paper and of
earlier work.

\section{Saturation effects and the stabilization of asters} 

So far, in our analysis, we have ignored the
effects of interactions between bound motors.
These motors, in their physical setting, have a
finite size and move on a one-dimensional track,
the microtubule. It is thus reasonable to expect
that interactions between bound motors should
become increasingly important at large motor
densities.  One can
account for these interactions by simply
requiring that more than one motor cannot occupy
the same volume in space at the same time. Such
``excluded volume'' interactions have been used
in earlier models for collective effects in
molecular motor transport\cite{menon}.

At the simplest level, accounting
for such interactions leads to a non-linear
(in general, also non-monotonic)
dependence of the motor current on the density.
This can be incorporated in our calculation
by choosing a motor current of the form
\be
{\bf J_{m_b}} = A m_b f(m_b) {\bf T}.
\ee
The function $f(m_b)$ should, in principle, be calculated from
an underlying microscopic model. We will circumvent this
necessity by simply assuming a convenient form purely
for illustrative purposes which
is (a) consistent with the requirement that $A$ (the
velocity) is density independent for small $m_b$
densities and (b) saturates for large $m_b$.
One such choice is 
\be
f(m_b) = 1 - \tanh(m_b/m_{sat}).
\ee
The value $m_{sat}$ limits the current of bound motors. When
$m_{sat} \gg m_b$ then $f(m_b) \rightarrow 1$ and we recover
results discussed earlier. When $m_{sat} \ll m_b $, the current
reduces as $f(m_b) \rightarrow 0$. In the aster case ({\bf T} 
= -$\hat r$), Eq.\ \ref{astercondition} becomes
\be
\partial_r m_f(r ) = - m_b(r) f(m_b).
\ee
Therefore, the free motors obey
\be
\partial_r^2 m_f + (\frac{1}{r} - \btof/f(m_b)) \partial_r m_f - \ftob m_f = 0.
\ee
This equation can now be solved as a boundary value problem 
using the methods described earlier.

\begin{figure*}
\begin{center}
\includegraphics[scale = 0.50]{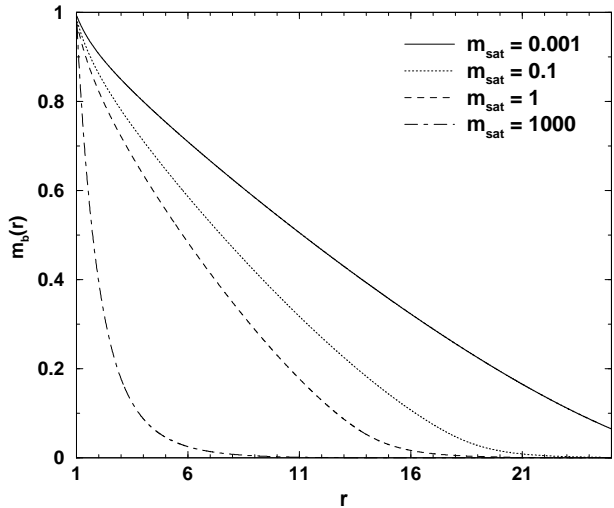}
\end{center}
\caption{\label{saturation}
Effect of saturation on the motor density profile. As the
effect of saturation is increased, the profile is smoother. The curves
are labeled from left to right as $m_{sat} = 1000,1.0,0.1,0.001$.
}
\end{figure*}

Figure 9 shows bound motor densities $m_b(r)$  as a
function of $r$ for different values of
$m_{sat}$. We have normalized each of the plots
to the same value at the origin. We choose
m$_{sat} = 0.001, 0.1, 1$ and $1000$. An
increased $m_{sat}$ implies smaller saturation
effects. The plots shown in Fig. 9 are normalized
to unity at $r = 1$.  Observe that reducing
m$_{sat}$ leads to far smoother variations of the
bound motor density.  Data for the associated
free motor densities are similar and are not
shown here.

\section{Relative stability of asters and vortices}

As we have seen in Section IV, single asters are
unstable to single vortices for large systems,
provided $S=0$. This is surprising because both
asters and vortices are equivalent topological
defects -- an aster can be transformed into a
vortex by rotating all ${\bf T}$  vectors through
$90^o$.  However, motor density profiles in aster
and vortex configurations are different.  It is
thus natural to seek an explanation for the
relative stability of asters and vortices at
$S=0$ which proceeds from this observation.

Lee and Kardar rationalize the stability of
vortices {\it vis. a vis}\/ asters in the
simulations in the following way: If
$\epsilon=1$, the right-hand-side of the tubule
equation derived by LK can be interpreted as the
functional derivative of a ``free energy''.  The
motor density profile plays the role of a
spatially modulated stiffness.  This enables the
calculation of the relative ``free energies'' of
aster and vortex states.  Smoother motor density
profiles lead to lower ``free energies''.  The
reduced energy of the vortex configuration, in
which motor density profiles decay
logarithmically, implies its increased stability
as compared to asters, for which motor density
profiles decay exponentially away from the core.

The competition between strain energies of asters
and vortices thus implies a system-size
dependence of ``energies'' and a cross-over
length scale at which the energies of the two are
comparable. For system sizes smaller than this
length, asters are favoured while vortices are
favoured for larger system sizes.
This simple argument rationalizes the relative
stability of aster and vortex states. We will use
it to explain an intriguing feature of our data
-- the presence of stable asters for fairly large
system sizes {\it e.g.} $L = 50$, in the $S=0$
limit.  At these scales and for equivalent
parameter values, the LK model would have yielded
a single vortex in steady state.  

We adapt the Lee-Kardar
argument to our model, using our analytic results
for bound motor density profiles in asters and
vortices. In Fig. 10, we show the {\em modulus}
of the difference in energy of a single vortex
and a single aster as a function of the system
size for our model and for the Lee-Kardar model.
We work at the same value of total motor density
$m$ in each case, setting $\epsilon=1$. Since the
magnitude of ``energy'' scales in the two models
differs, the behaviour of the energy difference
is most clearly seen using a logarithmic scale
for the $y-$axis. The point at which the vortex
energy is reduced below the aster energy appears
as the minimum in the curve.  From the position
of this minimum, we see that the crossover
length scale above which asters are unstable 
to vortices is smaller in the Lee-Kardar model 
as compared to our model.

\begin{figure*}
\begin{center}
\includegraphics[scale=0.5]{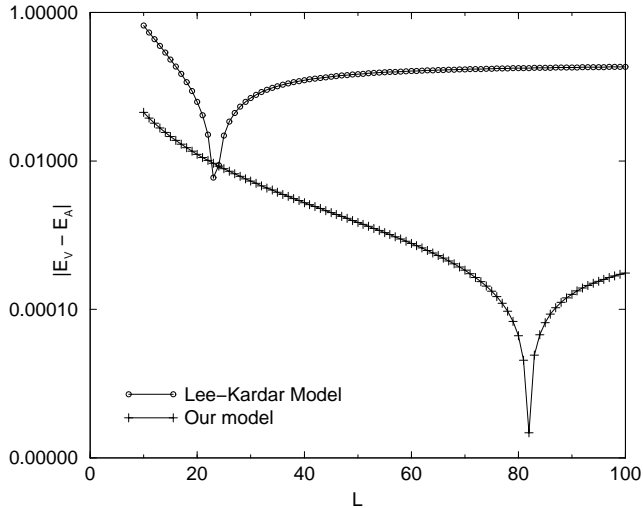}
\end{center}
\caption{\label{crossover}
A plot of the difference in energies of a
single vortex and a single aster versus the system size $L$.
The parameter values are $m=1,\epsilon=1$ and $S=0$. The
y-axis is plotted with a logarithmic scale.
The point at which the
energy of the vortex becomes lower than the energy of the aster shows up as
the minimum in this curve. In the Lee--Kardar model the length scale
at which the vortex becomes lower in energy is about four times smaller
than in our model.}
\end{figure*}

The fact that this crossover lengthscale lies
between $50$ and $100$ in our units for length 
rationalizes our observation that the stable
asters we obtain for $L=50$ are replaced by
stable vortices for $L=100$.  These results
support, and add further credence to, the
Lee-Kardar argument, at least in the $S=0$ case.
With finite $S$, of course, vortices are
disfavoured altogether and generating stable
aster structures is no longer an issue.

We now provide a qualitative explanation of three
basic features of motor and microtubule
arrangements at zero and non-zero $S$.  These are
(i) the absence of vortices at non-zero $S$, (ii)
the presence of a small number of large asters at
small $S$ {\it vs.} (iii) the presence of a large
number of small asters which increases as $S$ for larger  
$S$.  Our approach is approximate and qualitative, based
on ``free energy'' type arguments, analogous to 
those of Lee and Kardar.

Consider the limit in which $\epsilon = 1$ and
the $\nabla m$ term is obtained
through a functional differentiation of the
``free-energy'' term
\be
E_S  = \int d^2 x (\alpha\nabla \cdot {\bf T} + \frac{S}{\alpha}m_b)^2.
\ee
A functional derivative of $F$ with respect to the
{\bf T} field generates the terms
$\alpha^2 \nabla (\nabla \cdot {\bf T})$ and
$S\nabla m_b$. In the limit that
$\alpha \rightarrow 0$, we obtain the $\nabla m_b$\
term of our equation of motion. (We prefer to
work with a non-vanishing, although infinitesimal, 
value of $\alpha$ and to consider the effects of 
varying $S$, to avoid the singular behaviour which 
arises in the $\alpha=0$ limit.) Our introduction
of the term in $S$ was, in fact, motivated by
the observation that motors moving along two
initially parallel microtubule configurations,
act to bring them together, favouring a non-zero
divergence. This physics, though incorporated in the
simulations of N\'ed\'el\'ec {\it et al.}, is ignored
in the LK model.

The full ``free energy'' for the {\bf T} field is
\be
E  = \int d^2 x \left[-\frac{1}{2}\alpha |T|^2 
+ \frac{1}{4}\beta |T|^4 
+ \frac{1}{2}m_b(r)(\nabla {\bf T})^2 + 
\frac{1}{2}(\alpha\nabla\cdot{\bf T} + \frac{S}{\alpha}m_b)^2\right].
\ee
We fix the modulus of {\bf T}, thereby
eliminating the first two terms, and concentrate on
the relative energetics of states governed by
the last two terms of the equation.

For the Lee-Kardar model, with $S=0$, the
stability of vortices over asters can be
understood from simple comparisons of the
relative energies of vortices and asters. First
consider the case of a single aster. We choose a
simplified profile for bound motors,
assuming,
\bea
m_b &=&  A \hspace*{1cm}  r \leq \xi \nonumber \\
m_b &=&  0 \hspace*{1cm} r >  \xi.
\eea
This represents, although approximately, the fact
that microtubule densities are significant only
over distances of order $\xi$ from the core of
the aster.  The factor $A$ is a normalization
factor which ensures a fixed number of motors $N$
in a system of typical size $L$. We will work
with a fixed overall density of motors $\sim
N/L^2$ and examine the limit in which $L
\rightarrow \infty$.

In the $S=0$ limit, the energy $E_A$ of the aster 
is dictated  by
\be
E^0_A \sim\int_a^L d^2r~m_b(r)(\nabla {\bf T})^2 
\sim A \log(\xi/a)
\ee
where we cut the integral off at a lower limit $a$
corresponding to a microscopic coarse-graining
scale and have used the fact that 
$({\bf \nabla T})^2 = 1/r^2$ for both asters and 
vortices.
We now fix $A$ by normalization: since $\int_a^L
m(r)d^2r =  N = nL^2 \sim A\xi^2$, we obtain $E^0_A
\sim nL^2 \log(\xi/a)/\xi^2$. (Accounting for a more complex decay
of the bound motor density does not change this
result qualitatively.) Thus, for a system
of size $L$ ($L \gg \xi$), imposing a fixed {\em
density} of motor yields a quadratic increase of
the energy with system size $L$.

We now repeat the same argument with a single
vortex. Here, the relatively slower variation
$m_b(r) \sim$ C$_1$ + C$_2\log(r/\lambda$) yields
$E^0_v \sim \int_a^L (m_b(r)/r^2) r dr \sim
\log(L/a)$, upto further logarithmic corrections.
The prefactor is then {\em independent} of system
size, leading to an overall logarithmic increase
of energy with system size, a far weaker function
than the quadratic dependence on $L$ of the
aster. This implies that for sufficiently large
$L$, the energy of the isolated vortex falls
below that of a single aster, as first suggested
by Lee and Kardar.  in systems of sufficiently
large sizes.

How are these arguments modified at finite $S$? Since
simulations indicate that the state obtained
at finite positive $S$ is a ``lattice of asters'' with
the scale of the aster decreasing as $S$ is increased,
we will compare the energies of single vortex configurations 
with energies for an assembly of asters (``mini-asters'')
of typical size $\sigma$. We imagine that the $L\times L$
system is subdivided into $\sigma \times \sigma$ units
and place an aster in each one of these sub-units.
For a system of linear size $L$, the number density of such 
asters is $\phi$ .  For inward pointing asters, as obtained
in our calculation, the divergence ${\bf \nabla} \cdot {\bf T} 
= -1/r$. Consider first the term
\be
E_S \sim \int d^2 x \left[\alpha\nabla \cdot {\bf T} 
+ \frac{S}{\alpha} m_b(r)\right]^2.
\ee
Now mini-asters are (i) assumed small and (ii)
obey boundary conditions which differ from the
case of the single system-size 
spanning aster. Bound motor
profiles are then, in general, combinations of
the two solutions obtained earlier rather than the
single one which yielded the exponentially damped form
used in our earlier analysis. Also, the close
proximity of the many mini-asters formed indicates
a slower than exponential variation of the bound
motor densities about each one. Let us {\em assume} that motor
densities in such mini-asters adjust in order
that the following motor density
profile is obtained: 
$m_b(r) \sim \alpha^2/Sr (r \leq \sigma$) and 
$m_b(r) \sim 0 (r > \sigma$). This ensures that the
contribution of the term above is cancelled. 
Note that the bulk of the contribution to the energy of a 
mini-aster comes from the region $r \leq \xi$;
the contributions from larger regions is negligible provided 
$\xi \sim \sigma$.  

We first calculate the energy $E_{1mA}$ of a
single such mini-aster configuration.
This is obtained from the integral
\be
E_{1mA} = \int_a^L d^2r m_b(r) ({\bf \nabla T})^2 \sim
\frac{\alpha^2}{S}\int_a^\sigma dr \frac{1}{r^2} \sim \frac{\alpha^2}{Sa}
\ee
The number of motors in a single mini-aster is obtained from
\be
N_{1mA} \sim \int_a^\sigma m(r) d^2r \sim \frac{\alpha^2\sigma}{S}
\ee
The total energy $E^T_{mA}$ of the system of miniasters is then
\be
E^T_{mA} \sim \frac{\alpha^2}{Sa} \phi  L^2
\ee
The constraint $n = \phi N_{1mA}$ yields
\be
\phi \sim \frac{S}{\alpha^2 \sigma}
\ee
illustrating how the number of asters increases
as $S$ is increased. The total energy corresponding
to a fixed areal density of motors $n$ is then
easily derived as
\be
E^T_{mA} \sim \frac{nL^2}{\sigma a}
\ee

We now compare this with the energy of a single
vortex at non-zero $S$, computed assuming that the
motor density profiles are the same as in the case in
which $S$ was zero. This is obtained as
$E_v \sim S^2 n^2 L^2/\alpha^2$, yielding 
the ratio
\be
\frac{E_{V}}{E^T_{mA}} \sim \frac{n S^2 \sigma a}{\alpha^2}
\ee
We may also compare the energy of an assembly of mini-asters
with the energy of a single aster. Arguments similar 
to those above yield $E_A \sim n^2 S^2 L^2/\alpha^2$,
and thus
\be
\frac{E_{A}}{E^T_{mA}} \sim \frac{n S^2 \sigma a}{\alpha^2}
\ee

Note that both these arguments indicate that as
$\alpha$ is reduced towards zero (or $S$ is increased
from a non-zero value), both single asters and
single vortices are unstable to an array of
mini-asters. At infinitesimal $\alpha$, provided
$S$ is non-zero, our
results indicate that the number density of mini-
asters increases with $S$ (linearly in the simple
argument given here), with the size of each aster
decreasing in proportion.  The arguments given
here rest on the assumption that the motor
distribution in asters adjusts so as to minimize
the cost for the term involving $S$. 
No such adjustment can lower the energy of 
single vortex configurations.

\section{Conclusions}

This paper presents a hydrodynamic theory of
pattern formation in motor-microtubule mixtures,
accounting for the effects of confinement.  We
show that the influence of the boundaries on
pattern formation can be considerable, by
illustrating how either asters or vortices can be
formed depending on how the orientation of
microtubules is fixed at the boundary.  We have
explored the parameter space of $\epsilon, m$ and
$S$ systematically, describing the variety of
configurations obtained. We obtain density
distributions of free and bound motors
corresponding to the final microtubule
configurations. Such plots may be compared
directly to experiments. We have also compared
analytic predictions for motor density profiles
in isolated vortices and asters with simulation
data. Our results presented here complement 
our earlier work on pattern formation on 
much larger systems, in which the effects of
boundaries were minimal\cite{elsewhere}.

One technical improvement would be to account for
variations in the local {\em density} of
microtubule, as opposed to only their
orientation. We could then account for the
density dependence of quantities such as
$\ftob$.  More information from experiments,
performed in confined geometries using motors
with a range of different processivities would
also be useful in clarifying some of the issues
which relate to the simulations described here.
It would also be interesting to search for the
novel configurations we obtain here in different
regimes of parameter space, such as the ``flag'',
the distorted vortex and the ``outward aster''.

As we have emphasized here and in earlier
work\cite{elsewhere,physicascripta}, the set of
hydrodynamic equations we motivate and use allow for a
minimal yet complete description of the patterns formed in
mixtures of motors and microtubules. We have suggested
elsewhere\cite{physicascripta} that it may be useful to
think of spindle formation itself as a pattern
formation problem which can be modelled using continuum
hydrodynamical equations in a small number of variables.
Further work relating to this program is in progress.

We thank Y. Hatwalne, Madan Rao and G. Date for
useful and illuminating discussions. PBS thanks
IBM for providing computational facilities under 
a shared university research grant.


\begin{thebibliography}{99}
\bibitem{hydrodynamic} Since the experiments are performed in a
confined geometry, the otherwise long-ranged hydrodynamic 
interaction is screened because the boundaries of the system act
as a sink for momentum.  Neglecting the role played by 
hydrodynamics is therefore justified.
\bibitem{fail} We only fail to
obtain ``bundles'', a disordered phase obtained
at still higher densities of motors.
\bibitem{notephase} We find it convenient to
refer to the distinct states seen in
the experiments and our simulations as
``phases'', insofar as quantitative distinctions
can be made between them and they are well
contrasted from the point of view of the
experiments.  However such terminology is, strictly
speaking, inaccurate since we believe that sharp
qualitative distinctions between several of these
states, such as the disordered phase, the
aster-vortex mixture and the lattice of vortices
cannot be made -- a more generic name
would be a disordered/aster-vortex mixture
dominated by arrested states at low motor
density\cite{leekardar} in this regime.
\end{thebibliography}
\end{document}